\documentclass{amsart}
\usepackage{amssymb, latexsym} 
\newtheorem{theorem}{Theorem}

\begin{document}
\title{On the Structure of Dyck Languages}
\author{Rita Gitik}
\address{ Department of Mathematics \\ University of Michigan \\ Ann Arbor, MI, 48109}
\email{ritagtk@umich.edu}
\author{Eliyahu Rips}
\address{ Institute of Mathematics \\ Hebrew University, Jerusalem, 91904, Israel}
\date{\today}

\begin{abstract}
We prove that the closure of the one-sided Dyck language in a free monoid is a two-sided Dyck language.
\end{abstract}


\maketitle


\section{Introduction}

A Dyck language (cf. \cite{Be}, p.27) consists of "well-formed" words over a finite number of pairs
of parentheses. The restricted or one-sided Dyck languages $D_n^{'*}, n \geqslant 1$ are formed of the words over $n$ pairs of 
parentheses which are "correct" in the usual sense, i.e. in each pair of canceling brackets 
an opening bracket precedes the closing bracket.  Thus $( [ ( ) ( ) ] \{  \} [ ] ) ( )$ is a word in $D_3^{'*}$.

For the unrestricted or two-sided Dyck languages $D^*_n, n \geqslant 1$, the interpretation of the parentheses
is different. Two parentheses of the same type are considered as formal
inverses for each other. A word is considered as "correct" if and only if successive
deletion of pairs of associated parentheses of the form $()$ and of the form $)($ yields the
empty word. Thus $)()(][  )($ is a word in $D_2^*$.
 
Note that the word $)($ is a word in the two-sided Dyck language, but not in the one-sided Dyck language. 

Let $D^*_2$ be a two-sided (unrestricted) Dyck language on  two pairs
of letters $\{ a, \bar{a} \}$ and $\{ b, \bar{b} \}$. Denote by $D^{'*}_2$ the corresponding one-sided (restricted)
Dyck language.

Let $M= \{ a, \bar{a}, b, \bar{b} \}^*$ be the free monoid on $\{ a, \bar{a}, b, \bar{b} \}$ and let $F$ be the free group on the free generators
$\{ a, \bar{a}, b, \bar{b} \}$. We endow $F$ with the profinite topology in which all the subgroups of finite index in $F$ are the open neighborhoods of $1_F$. We consider the topology on $M$ induced by the embedding $M \rightarrow F$.

The main result of this paper is the following theorem.

\begin{theorem}
The closure of $D^{'*}_2$ in $M$ is $D^*_2$.
\end{theorem}

\section{Proof of Theorem 1}

Let $F_2$ be the free group on the generators $x$ and $y$. Let $\phi : F \rightarrow F_2$ be the homomorphism given by 
$\phi(a)=x, \phi(\bar{a})=x^{-1}, \phi(b)=y$, and $\phi(\bar{b})=y^{-1}$. Then $Ker(\phi) \cap M = D^*_2$ because $Ker(\phi)$ consists of words in $F$ which reduce to $1_F$ when we delete the subwords $a \bar{a}, \bar{a} a, b \bar{b}, \bar{b} b, a^{-1} (\bar{a})^{-1}, (\bar{a})^{-1} a^{-1},
 b^{-1}(\bar{b})^{-1}$, and $(\bar{b})^{-1} b^{-1}$. All such words constitute the two-sided Dyck language $D^*_2$.
 
The homomorphism $\phi$ is continuous in the profinite topologies on $F$ and $F_2$ because a preimage of a subgroup of finite index in $F_2$ is a subgroup of finite index in $F$. Therefore, $Ker(\phi)$ is closed in $F$ and $Ker(\phi) \cap M =D^*_2$ is closed in the induced topology on $M$.
Thus, the closure of $D^{'*}_2 $ in $M$ is a subset of $D^*_2$.
 
On the other hand, let $U$ be a normal subgroup of finite index in $F$. There exists $n$ such that $(aU)^n =( \bar{a} U)^n=(bU)^n =(\bar{b}U)^n=U$.
It follows that $a^{n-1}\equiv a^{-1}(mod U), (\bar{a})^{n-1} \equiv (\bar{a})^{-1}(mod U), b^{n-1} \equiv b^{-1} (mod U), (\bar{b})^{n-1} \equiv (\bar{b})^{-1} (mod U)$.
Hence $\bar{a} a = ({(\bar{a}a)^{-1})}^{-1} = {(a^{-1} (\bar{a})^{-1})}^{-1} \equiv 
{(a^{n-1} (\bar{a})^{n-1})}^{-1}(mod U)$ and $\bar{b} b = ({(\bar{b}b)^{-1})}^{-1} = {(b^{-1} (\bar{b})^{-1})}^{-1} \equiv 
{(b^{n-1} (\bar{b})^{n-1})}^{-1}(mod U)$.
  
Therefore $D^*_2 \subseteq \langle D^{'*}_2 \rangle U$, where $\langle D^{'*}_2 \rangle $ is the subgroup of $F$ generated by $ D^{'*}_2 $ in $F$.
It follows that the closure of $D^*_2$ in $F$ is a subset of the closure of $\langle D^{'*}_2 \rangle $ in $F$.
  
So ${\overline{D^*_2}}^M = {\overline{ D^{'*}_2}}^F \cap M = \overline{ D^{'*}_2}^M =  D^{'*}_2 $.
Hence, ${\overline{ D^{'*}_2}}^M = {\overline{ D^*_2}}^M = D^*_2$, as required.
  
\section{Acknowledgment}

The first author would like to thank the Albert Einstein Institute of Mathematics
of the Hebrew University for generous support.

\end{document}